%% file: main.tex
\documentclass{vgtc}                          

\usepackage{mathptmx}
\usepackage{graphicx}
\usepackage{times}
\usepackage{marginnote}
\usepackage{amsmath}
\usepackage{color}
\usepackage[normalem]{ulem}
\usepackage{hyperref}
\usepackage{footnote}

\onlineid{0}

\vgtccategory{Research}
\vgtcinsertpkg

\let\oldsqrt\sqrt
\def\sqrt{\mathpalette\DHLhksqrt}
\def\DHLhksqrt#1#2{%
\setbox0=\hbox{$#1\oldsqrt{#2\,}$}\dimen0=\ht0
\advance\dimen0-0.2\ht0
\setbox2=\hbox{\vrule height\ht0 depth -\dimen0}%
{\box0\lower0.4pt\box2}}

\title{Learning Style Similarity for Searching Infographics}

\author{Babak Saleh\thanks{This work has been done while Babak Saleh was at Adobe research.} \\ 
 \scriptsize\href{mailto:babaks@cs.rutgers.edu}{Rutgers University}%
\and  Mira Dontcheva \\ 
 \scriptsize\href{mailto:mirad@adobe.com}{Adobe Research}%
\and Aaron Hertzmann \\ 
 \scriptsize\href{mailto:hertzman@adobe.com}{Adobe Research}%
\and Zhicheng Liu \\ 
 \scriptsize\href{mailto:eoli@adobe.com}{Adobe Research}}%

\teaser{
\includegraphics[width= \textwidth]{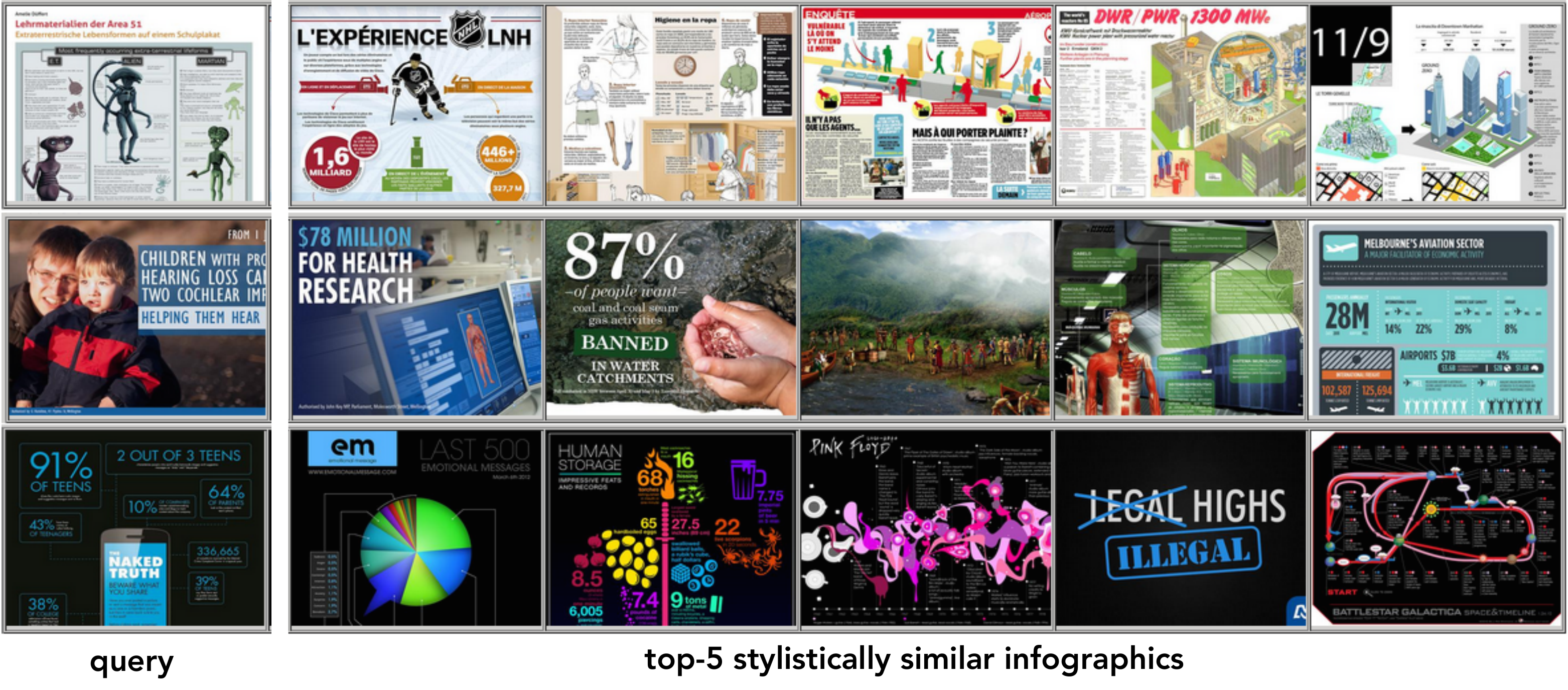}
\caption{Infographics combine text, charts and images. We consider the problem of learning style similarity for infographics to enable style-based search across a repository of infographics. For a query infographic (left), our approach returns the most stylistically similar infographics. This figure shows three example queries, one in each row.}
\label{fig:motivation}
\vspace{-0.2in}
}

\abstract{Infographics are complex graphic designs integrating text, images, charts and sketches. Despite the increasing popularity of infographics and the rapid growth of online design portfolios, little research investigates how we can take advantage of these design resources. In this paper we present a method for measuring the style similarity between infographics. Based on human perception data collected from crowdsourced experiments, we use computer vision and machine learning algorithms to learn a style similarity metric for infographic designs. We evaluate different visual features and learning algorithms and find that a combination of color histograms and Histograms-of-Gradients (HoG) features is most effective in characterizing the style of infographics. We demonstrate our similarity metric on a preliminary image retrieval test.}
\begin{document}


\maketitle

\section{Introduction}
\label{sec:intro}
\input{intro}

\section{Related work}
\label{sec:relwork}
\input{relwork}

\section{Overview}
\label{sec:overview}
\input{overview}

\section{Crowdsourcing similarity data}
\input{turk}

\section{Modeling similarity in infographics}
\label{sec:model}
\input{model}

\section{Results}
\input{exp}

\section{Conclusions and future work}
In this paper we investigate the problem of style similarity for infographics. Based on human perception of similarity gathered through crowdsourcing, we model stylistic similarity for infographics using low-level visual features. We evaluated a number of different features and found that a combination of color histograms and HOG features performs the best in predicting similarity. We applied this method to demonstrate a style-based search engine for infographics. As part of our work, we created a novel dataset of 19,594 infographics, which we plan to share with the community. 

In future work we plan to continue our investigations in this space. First, we want to evaluate our work with infographics designers. Feedback from those trying to use the a search interface for infographics will give us a sense for whether the current accuracy is sufficient and what is most important to designers. We also plan to increase the scale of our dataset and collect more human subject data. 

Second, we are interested in supporting search through stylistic keywords. Currently, our method does not support searching for \emph{minimalist} designs,  infographics with \emph{a three-column layout}, or infographics that show \emph{timelines}. To support this type of keyword search, we need to develop new visual features specifically tuned to infographics. In particular, we want to explore adding chart classification as inspired by the ReVision system~\cite{revision}. We did some initial tests in applying the existing ReVision classifiers for charts, such as bar charts or line graphs, but found that we must first develop methods for locating these types of charts in the infographic before we can classify them. 

Third, we would like to build new features that are especially designed for infographics. This needs human expertise in terms of designers knowledge about making infographics and their preferences for finding similarity in infographics.


Finally, we are interested in exploring style retargeting, as has been demonstrated in the context of web design~\cite{bricolage} and more traditional data visualization~\cite{revision}. Since we do not have the structural DOM information available, we will have to rely on computer vision to do data extraction. Savva et al.~\cite{revision} show that robust data extraction is possible, but since infographics designs are more complex and include additional visual elements beyond charts and graphs, more sophisticated analysis methods will be necessary.
 


\bibliographystyle{abbrv}
\bibliography{main}
\end{document}

%% file: intro.tex
Infographics are increasingly used to tell visual stories about data, by combining text, charts, and images. However, it remains difficult to design effective infographics. 
As in any kind of graphic design, an important step in the design process is to examine existing resources for inspiration. While today's search engines allow designers to search through keywords, no tools exist for exploring infographic designs by style. Ideally, designers should be able to browse through designs using keywords such as ``minimalist'' or ``retro'' or perform search by example to find similar or drastically different example designs.

Previous work for searching graphic design has primarily focused on domains where
vector representations of the exemplars are available, e.g., HTML or DOM hierarchies~\cite{dtour, webzeitgeist}.  
However, most infographics on the web are available only as bitmap images, which offer no direct access to the shapes, colors, textures, text, images, charts, and underlying data present in each infographic. Stylistic analysis and search are thus especially challenging.

In the absence of the underlying data for a design, how well can we search for style, given only the pixel information inside an infographic?  Qualitatively, we observe that there is a large variety of design styles for infographics --- much more than in typical web designs --- making the problem of defining similarity for infographics more difficult. Because infographics include a wider variety of content elements, such as blocks of text and charts, similarity sits in a higher-dimensional space and thus requires more data and a wider variety of image features.  
Our ultimate goal in this research is to study similarity of graphic designs in general. However, in order to focus the problem, we analyze infographics as a more-constrained special case that is interesting and challenging in its own right.

In this work, we compute style similarity between infographics based solely on low-level visual features, inspired by their success in the computer vision literature.  We collect a dataset of 19,594 infographics from the web, along with crowdsourced similarity ratings for a subset of this collection from human subjects on Amazon Mechanical Turk.  This dataset allows us to learn a similarity model from these examples.  We experiment with several types of visual features for measuring similarity, and use a held-out subset of the ground truth data for evaluation.  We find that a combination of color histograms and Histograms-of-Gradients (HoG) features works best for learning similarity, in comparison to the other, primarily low-level, visual features that we tested. Our work is exploratory, we leave comprehensive study on the use of high-level features for future. We demonstrate the method by showing search-by-similarity results on the full dataset (Figure 1).

%% file: relwork.tex
Our work lies at the intersection of data visualization, graphic design, and learning similarity based on human judgments. To the best of our knowledge, this paper is the first to explore style similarity for infographics. 

Researchers have conducted experiments to understand what makes infographics effective and have found that visual embellishments including recognizable cartoons and images, elements that are common to infographics, enhance data presentation and memorability~\cite{bateman2010useful, borkin2013makes} . 
There also has been some previous work on applying computer vision algorithms to data visualizations. Prasad et al.~\cite{prasad} categorize simple charts types, namely, bar charts, curve plots, pie charts, scatter plots and surface plots based on low-level visual features of bitmap images (HoG and SIFT).
Savva et al.~\cite{revision}  classify a number of chart types, including bar charts, pie charts and line graphs, extract the data from the visualization, and suggest improved design layouts to visualize the same data in a better way. 
Both of these works require each input to comprise only a single chart type, and both works attempt to factor out the effects of style. In contrast, we consider infographics that may comprise complex arrangements of elements, and we focus on comparing style without detailed parsing of the inputs.



Several previous systems have explored search-by-style for web design.
Ritchie et al.~\cite{dtour} propose a search interface for web design that supports style-based search based on a set of style features including layout, color, text, and images.  Kumar et al. search for web designs based on style in a large dataset of 100,000 webpages~\cite{webzeitgeist}, and demonstrate style retargeting across website designs by crowdsourcing the mapping between designs~\cite{bricolage}. Chaudhuri et al.~\cite{attribit} predict stylistic attributes of web designs.
In each of these methods, it is assumed that the full HTML/DOM hierarchy is available and there is a limited space of possible layouts.  In contrast, we focus on infographics, for which vector data is rarely available, and thus, we must begin from bitmap input.

Our work is inspired by methods for style-based search in line drawings~\cite{hurtut2011artistic}, illustrations~\cite{elena}, and fonts~\cite{peter}. Unlike previous work, we focus on infographics, which include heterogenous elements arranged in complex layouts.

%% file: overview.tex
Our goal is to determine the stylistic similarity between any two infographics. Given two bitmap images of infographics, our model returns a numerical score evaluating the stylistic similarity between the two infographics. Our approach for learning and modeling similarity is similar to~\cite{elena} and~\cite{peter} and is summarized in Section \ref{sec:model}. 
To train our similarity model, we crowdsourced similarity ratings for a subset of our dataset. We collected similarity ratings from human subjects using Amazon Mechanical Turk. We demonstrate the method through a search-by-example application (see Figure \ref{fig:search}).

\section{Dataset of infographics}

To the best of our knowledge there is no established dataset of infographics. Thus, we created a dataset from the Flickr website. We chose Flick because it has a large collection of infographics and we could easily check for Creative Commons licensing. 
We gathered this collection by querying with the keyword ``infographic" and downloading Creative Common images with high or medium resolution as defined by Flickr.  We pruned  images that were photographs by inspecting the XMP metadata of each downloaded image. Additionally, we manually inspected the collection and removed images that appeared to be photographs or drawings. In total we pruned 2,810 images resulting in a dataset of 19,594 infographics (9,088 high resolution and 10,506 medium resolution). High resolution images have width 1024px and height 768px. Medium resolution images include images with resolutions of 800x600, 640x480 and 500x375.

%% file: turk.tex
\label{sec:turk}


\begin{figure}
\includegraphics[width= \columnwidth]{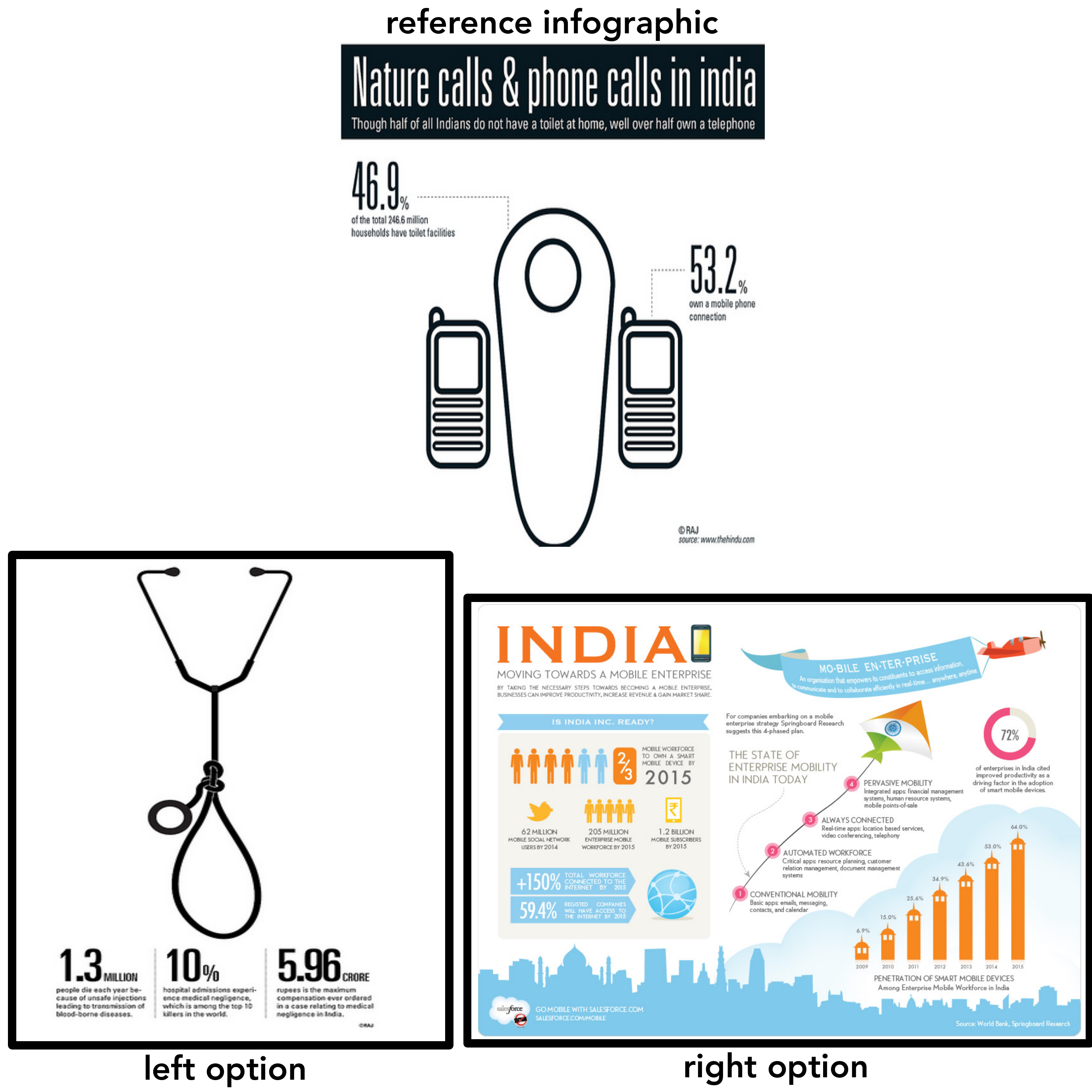}
\caption{The design of our Amazon Mechanical Turk experiment for collecting similarity data. Despite the fact that both the reference infographic (top) and right option are about a similar topic (cellphone usage in India),  the left option is more \textit{stylistically} similar to the reference infographic.}
\label{fig:turk}
\end{figure}

We designed a human subject experiment to measure the similarity between two infographics. In our experiment, human subjects were asked to compare two infographics to a reference infographic and select the one that is \emph{stylistically} more similar (Figure \ref{fig:turk}). Since asking experts to annotate this large scale data set is expensive, we used Amazon Mechanical Turk to run our experiment. It is possible that professional designers may give more accurate style and similarity annotations. In this platform, each experiment is considered a human intelligence task (HIT).  Our HIT included an introduction session with three examples that taught the user the purpose of the experiment and explained the meaning of stylistically similar infographics. Raters were instructed to focus on stylistic similarity and to ignore content semantics, with training tasks meant to illustrate this. The example shown in Figure \ref{fig:turk} was used in the training. The infographic in the bottom left is stylistically similar but is on a different topic. The infographic in the bottom right is stylistically different but is on the same topic (cell phone usage in India). The correct answer for this example is the left option. 

%

After the training session, users were asked to answer 20 questions each showing a different set of images. We used three control questions to verify the quality of answers. The control questions were easy questions with obvious correct answers. If the user missed two of the control questions, the HIT was rejected and the user was banned from additional experiments.  
Each HIT was completed by at least 9 people, and we paid \$0.3 per HIT to each user. 

The final version of our infographics dataset has 19,594 bitmap images. Since it is practically impossible to gather similarity data for all of the images, we randomly selected 2,082 images from the 9,088 high resolution images for annotation by human subjects. From the 2,082 images we created 847 \emph{triplets}. A triplet includes the three images we showed together in each question, the reference image and the two options. Because of the random image selection process, some images were used in more than one triplet. 
  

\subsection{Results}


We collected  8,454 ratings across 847 triplets. On average, users took 3 minutes to complete the entire task. After removing  answers from users who did not pass the control questions, our dataset included at least 9 responses for each triplet. We consider the majority's choice as the correct answer for each comparison. Table~\ref{tab:accuracy_acc} \& Table~\ref{tab:accuracy_exc} show how accuracy is affected by changing the threshold for which triplets are considered correct. For example, on the subset of triplets for which 60\% or more raters agree, 79.59\% of raters give the correct answer (Table~\ref{tab:accuracy_acc}). While for subset of triplets which 60-70\% raters agree, 60.21\% of raters give the correct answer (Table~\ref{tab:accuracy_exc}). Participants in our experiment indicated United States, India, Italy and Australia as their nationality, and 53\% of them reported being female. The dataset and annotations are provided on the accompanying website.

		
\begin{table}
\small
	    \begin{tabular}{ | l | l | l | l | l | l | l |}
	    \hline
		Threshold(\%) & 50 & 60 &70 &80 &90 & 100 \\
		\hline
			\hline
		 Responses \footnotemark & 8454	&7549	& 5840	&4402	&2985	&1515\\ \hline
		 			Triplets \footnotemark  & 847 & 756 & 585 & 441 & 299 & 152\\ \hline
		Accuracy(\%) \footnotemark  & 76.45	&79.59	&85.31	&90.28	&95.08	&100\\ \hline
		
	    \end{tabular}
	\caption{Analysis of triplet annotations based on the user agreement.} 
	\label{tab:accuracy_acc}
	\end{table}
%

\begin{table}
\small
	    \begin{tabular}{ | l | l | l | l | l | l | l |}
	    \hline
		Threshold(\%) & 50-60 & 60-70 & 70-80 & 80-90 & 90-100 & 100 \\
		\hline
			\hline
		 Responses   & 905	& 1709	& 1438	& 1417	&1470	&1515\\ \hline
			Triplets & 91 & 171 & 144 & 142 & 147 & 152\\ \hline
		Accuracy(\%) & 50.28	& 60.21	& 70.1	& 80.17	& 90 	& 100\\ \hline
		
	    \end{tabular}
	\caption{Analysis of triplet annotations based on the user agreement.} 
	\label{tab:accuracy_exc}
	\end{table}
	\footnotetext{i.e., what percentage of the time, Turkers are correct on these images}
	\addtocounter{footnote}{-1}
	\footnotetext{i.e., number of triplets in this category}
 	\addtocounter{footnote}{-1}
	\footnotetext{i.e., number of responses in this category}
 


In order to measure the agreement between annotators in our experiment, we computed  the consistency of each responder. For each triplet we took the majority's opinion as the ground truth, and we counted the number of times each user picked the ground truth option across all assigned triplets. On average 76.5\% of users picked the ground truth (we call this measure ``Oracle" — which always picks the option selected by the majority of annotators — in Table~\ref{table:measure}). This number intuitively shows how consistent is the provided annotation. Similar measurement has been done in related work~\cite{elena,peter} and we concluded that our data has a similar consistency in responses from annotators.
%
%

%% file: model.tex

A key hypothesis of our work is that style can be described with low-level visual features.  For example, a ``busy'' design with lots of elements and textural detail has a much higher entropy gradient histogram than a very ``minimal'' design. Likewise, color factors heavily into the style of design. Bright colors evoke a different feeling than light or subdued colors.

\subsection{Features}
\label{subsec:imgRep}



We explored a variety of visual features inspired by previous work in computer vision. We primarily focused on low-level image features that have been successfully used for object recognition in photographs. However, we also tested visual features that have been found to be successful in encoding similarity in clip art~\cite{elena} and object classifier approaches, such as PiCoDes, which identify specific objects in natural photographs (e.g. faces, wheels and chairs).


The low-level visual features we explored include GIST, Histogram-of-Gradients (HoG), Local Binary Patterns (LBP) and histograms of color and luminance. 
\begin{itemize}
\item GIST~\cite{gist} provides a low-dimensional representation that represents the dominant spatial structure of an image.  
\item The Histogram-of-Gradients (HoG) feature~\cite{hog} is computed as a histogram of image derivatives with bins across both spatial and orientation axes. It has been shown to be very effective for object detection. 
\item The Local Binary Pattern (LBP) features~\cite{lbp} are designed to represent types of texture in a manner that is invariant to color or shading variations. LBP has been used for texture classification and object categorization.
\item Finally, color and luminance of the image are important for any judgment about the style of the infographics. We include histograms of colors and luminance as features.
\end{itemize}

\subsection{Feature implementation}

In order to have fixed-size feature vectors for images, prior to feature extraction we scaled landscape designs to a maximum height of 450 pixels and portrait designs to a maxim width of 360 pixels.  We also cropped each design to a window of 450px x 360px. We chose these dimensions based on a statistical analysis of our repository. 

To calculate GIST features we used the original implementation~\cite{gist}, and to implement HoG and LBP, we used the VLFeat toolbox~\cite{vlfeat}. We extracted HoG features with cell size 16 to capture finer details in the infographic and with cell size of 32 to capture information at a coarser level. 
To make learning over these feature vectors tractable (i.e. finish computing in under a day), we used Principal Component Analysis (PCA) to lower the dimensionality of GIST, HoG-16 and HoG-32 vectors to 230 dimensions.

We calculated color and luminance histograms manually. We set 10 bins for each color and luminance channel resulting in a 30 dimensional feature vector for color histogram and a 10 dimensional vector for the histogram of luminance. For exploring combinations of features, we applied PCA on each feature type separately and concatenate the output vectors to make the final feature vectors.





\

\subsection{Learning algorithm}
\label{subsec:learning}

We now describe the algorithm we use for learning style similarity between infographics. Our approach is an instance of metric learning \cite{kulis2012metric,Tamuz} based on methods used previously for fonts \cite{peter} and clip art \cite{elena}.

Given two infographics $X$ and $Y$, we compute their feature vectors $\textbf{f}_X$ and $\textbf{f}_Y$. The weighted distance between them is then:
\begin{equation}
D(X, Y) = \sqrt{(\textbf{f}_X - \textbf{f}_Y)^{T} \textbf{W} (\textbf{f}_X - \textbf{f}_Y)}
\label{eq1}
\end{equation} 
where $\textbf{W}$ is a diagonal matrix that weights the feature vector dimensions. 

 Given the crowdsourced data (Section \ref{sec:turk}), our goal is to learn the weights on the diagonal of $\textbf{W}$.  We model the response data as follows.  Suppose a human rater is shown a reference infographic $A$, and asked whether infographic $B$ or $C$ is stylistically more similar to $A$. We model the probability that the rater answers that $B$ is more similar to $A$ as a sigmoidal function of the pairwise distances:
\begin{equation}
P^{A}_{BC}=\frac{1}{1+\exp(D(A,B)-D(A,C))}
\label{eq2}
\end{equation}
In other words, when the distance between $A$ and $B$ is much smaller than the distance between $A$ and $C$, then the rater has very high probability of picking $B$. When the probabilities are nearly the same, the rater's response is nearly random.
The goal of learning is to estimate $\textbf{W}$ to most accurately predict (fit) the human ratings. 

As our feature vector has high dimensionality, we also regularize the weights with a Laplacian prior: $P(\textbf{W})\propto \exp(-\lambda \|\mathrm{diag}({\bf W})\|_1)$, with weight $\lambda$.  This prior is known to act as a sparsifying prior, potentially eliminating unnecessary feature dimensions.

As we explained in section~\ref{sec:turk}, we first filter annotations by removing the inconsistent responses. The final label of each triplet (0/1) is based on the label of the option that the majority of users picked for the reference.  Given all training triplets $\mathcal{D}$, learning is performed by Maximum A Posteriori (MAP) estimation, which entails minimizing the following objective function:
\begin{equation}
-\sum_{\mathcal{D}} \log P^{A}_{BC} + \lambda \|\mathrm{diag}({\bf W})\|_{1}
\label{eq3}
\end{equation}

As this optimization is non-linear and unconstrained, we optimize this function using L-BFGS~\cite{lbfgs}. Although the solver can enforce non-negative weights, we did not find this to be necessary as it produced non-negative weights explicitly using bounds constraints. We determine $\lambda$ by five-fold cross validation on the training set.  We found that the best results are given by setting $\lambda$ to 1.  We trained the model on 600 out of the 847 crowdsourced triplets and tested using the remaining triplets. On average learning took an hour on a desktop with a 3.7 GHz Intel Xenon quad core processor and 16 GB of RAM. 	

%% file: exp.tex
\label{sec:exp}

	\begin{table}
	\begin{center}
	    \begin{tabular}{ | l | l | l |}
	    \hline
		Approach & Dimensions & Accuracy(\%) \\
		\hline
			\hline
		GIST & 230 & 52.35\\ \hline
		LBP & 230 & 51.80\\ \hline
		HoG-16 & 230 & 57.65\\ \hline
		HoG-32 & 230 & 53.80\\ \hline
		Color histogram & 30 & 62.94\\ \hline		
		Luminance histogram & 10 & 40.83\\ \hline		
		Color histogram + GIST & 230 & 54.71\\ \hline
		Color histogram + LBP & 230 & 61.18\\ \hline
		Color histogram + HoG-16 & 230 & \textbf{71.83}\\ \hline
		Color histogram + HoG-32 & 230 & 59.13\\ \hline
		Similarity in clipart~\cite{elena} & 169 & 55.88\\ \hline
		PiCoDes& 230 &60.56\\ \hline\hline
		Baseline (no learning) & 230 & 59.92 \\ \hline 
		Oracle & & 76.45 \\ \hline
		
	    \end{tabular}
	\end{center}
	\caption{Quantitative comparison of different features for the task of similarity prediction} 
	\label{table:measure}
	\vspace{-0.25in}
	\end{table}

	To evaluate the different visual features, we compute accuracy through the percentage of correctly-predicted triplets. 	
	Table~\ref{table:measure} shows that color histograms perform remarkably well. Color works better than all other low-level visual features. It also performs better than the more sophisticated approaches, such as similarity features for clip art and higher-level object-classifier methods. Surprisingly, combining features does not always work better. While we achieve highest accuracy by combining color histogram and HoG-16 at 71.83\%, adding GIST, LBP, or HoG-32 does not improve accuracy and sometimes even lowers it.  We suspect that GIST brings down accuracy, because it overwhelms the color histogram features and itself includes color-based elements. 
HoG and LBP features are not correlated with color (designed to be color invariant), and thus we would expect them to complement the color histogram features. And indeed the small window size of HoG-16 features as compared to HoG-32 leads them to capture details at the right level.  With LBP we used a window size of 3, which was too small and captured noise. Additionally, unlike HoG, LBP also does not have any normalization across neighboring windows. In the end, color histograms and HoG-16 features perform the best. HoG-16 captures aspects of layout, density, and orientations, whereas color is crucial to style. 
	
	At 71.83\% this combination of features does almost as well as the oracle. As described in Section 5.1, ``oracle" refers to the best-possible performance based on our analysis of human subject data. Again, on average 76.5\% of users picked the ground truth infographic determined through majority vote. For additional comparison, we also define a baseline as the euclidean distance between feature vectors for the highest performing combination of features (color histogram and HoG-16).


\begin{figure}[t!]
\includegraphics[width=\columnwidth]{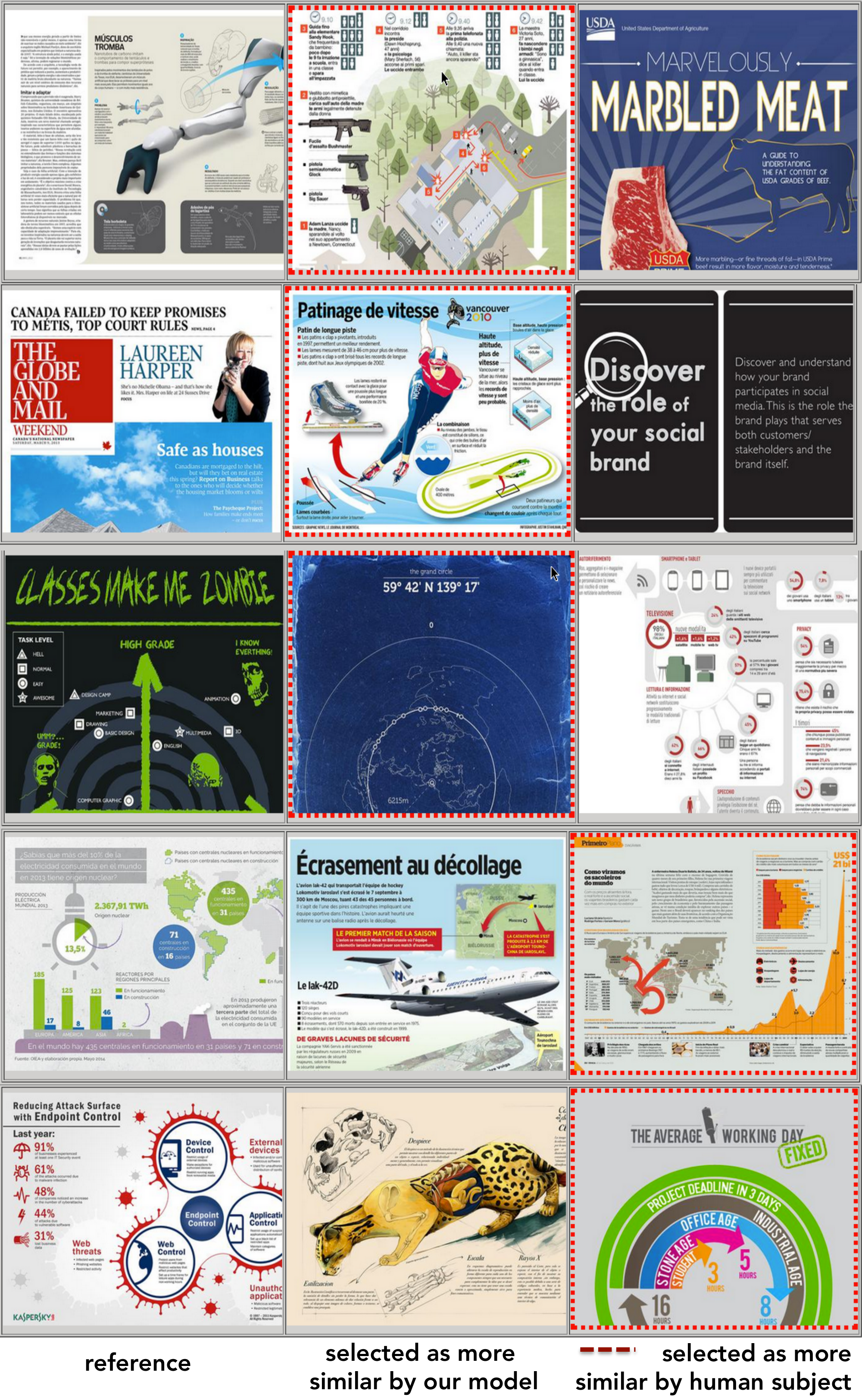}
\caption{Qualitative evaluation of the learned metric for the task similarity prediction. The first three rows show success cases, and the last two rows show failures of our model in predicting the more similar pair. In each row, the image on the left is the reference image. The middle image is the option that the model predicts to be more stylistically similar. The red dotted bounding box shows the human preference (what the majority of users picked). }
\label{fig:qual}
	\vspace{-0.25in}

\end{figure}





	Figure~\ref{fig:qual} shows some qualitative results. Each row shows a triplet. The first image is the reference image. The image in the middle is the one predicted by the model to be more similar. The image with the red dotted outline is the image selected by majority of human subjects as more similar (ground truth).

	\begin{figure*}[th!]
	\includegraphics[width=\textwidth]{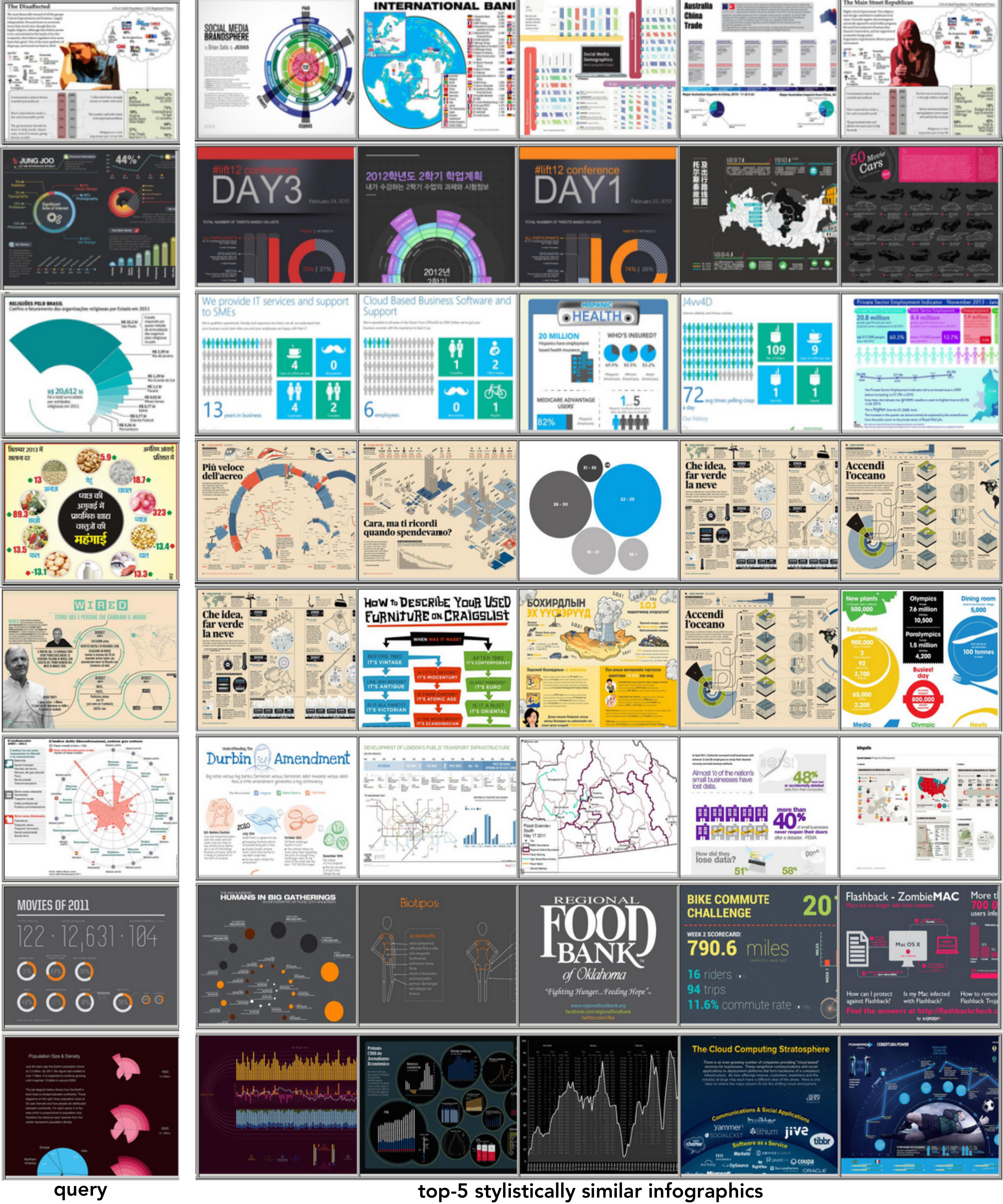}
	\caption{Sample results from our prototype for a search engine for infographics. In each row, we see a different example.  For a query design (left), we find the most similar infographics using stylistic similarity (right).}
	\label{fig:search}
	\end{figure*}

\section{Search engine for infographics}
	
Accurate similarity prediction allows us to build search engines that enable stylistic search over repositories of graphic designs. We have implemented an image retrieval algorithm that returns stylistically similar images to a given query image.
Figure~\ref{fig:search} shows sample results for our prototype search engine. Each row represents a different query. Our prototype retrieves the most stylistically similar designs for each query. We present top-5 retrieved infographics to show the consistency among the results. 

%